\documentclass[conference]{IEEEtran}
\IEEEoverridecommandlockouts
\usepackage{cite}
\usepackage{amsmath,amssymb,amsfonts}
\usepackage{algorithmic}
\usepackage{graphicx}
\usepackage{textcomp}
\usepackage{xcolor}
\usepackage{subfig}
\usepackage{booktabs}
\usepackage{cite}
\usepackage{tikz}

\def\BibTeX{{\rm B\kern-.05em{\sc i\kern-.025em b}\kern-.08em
    T\kern-.1667em\lower.7ex\hbox{E}\kern-.125emX}}
\begin{document}

\title{Learning to Maximize Energy Efficiency in 6G in-X Subnetworks

}

\author{\IEEEauthorblockN{Ramoni Adeogun}
\IEEEauthorblockA{\textit{Department of Electronic Systems} \\
\textit{Aalborg University, Aalborg, Denmark}\\
Email: ra@es.aau.dk}
}

\maketitle

\begin{abstract}
This paper investigates energy-efficient power control in 6G in-X subnetworks. We consider a graph neural network (GNN) framework that captures inter-subnetwork interference and the underlying wireless topology to optimize transmit powers. Three energy efficiency (EE) formulations are studied: (i) network-centric, which maximizes total network energy efficiency; (ii) subnetwork-centric, which maximizes the average energy efficiency per subnetwork; and (iii) a multi-objective approach, which balances energy efficiency and sum-rate performance. Extensive simulations in industrial factory settings with 3GPP channel models demonstrate that the GNN effectively learns interference-aware power allocation policies, significantly outperforming maximum power transmission and existing GNN based power control solution. Results showed network EE gains of up to 1341\%, average per-device EE improvements of up to 1302\%, and sum-rate enhancements up to 24.7\% relative to a maximum transmit power policy, depending on the chosen optimization formulation and trade-off settings.

%Extensive simulations demonstrate that the GNN-based solutions effectively learn interference-aware power allocation policies, achieving significant energy efficiency gains over benchmark strategies while maintaining high sum-rate performance. The comparative analysis provides insight into the trade-offs between global network performance and per-subnetwork efficiency, offering a flexible framework for energy-aware resource allocation in future 6G networks.
\end{abstract}


% \section{Introduction}
% The sixth generation (6G) of wireless communications is expected to support applications with extreme requirements in throughput, latency, reliability, and device density \cite{6Gpaper, 6Gpaper2}. Traditional macro- and small-cell deployments will be inadequate to meet these demands in ubiquitous, heterogeneous environments. A promising paradigm to address this challenge is the concept of in-X subnetworks — short-range, low-power radio cells embedded within physical entities such as a production module, a robot, a vehicle, a house, or even a human body \cite{Adeogun2020,Berardinelli2021}. These subnetworks would provide capillary wireless coverage, enabling life-critical or mission-critical communications that historically relied on wired connectivity. 
\section{Introduction}

The sixth generation (6G) of wireless communications is expected to support applications with extreme requirements in throughput, latency, reliability, and device density \cite{akhtar2020shift,saad2019vision}.
Traditional macro- and small-cell deployments will be inadequate to meet these demands in ubiquitous, heterogeneous environments. A promising paradigm to address this challenge is the concept of in-X subnetworks — short-range, low-power radio cells embedded within physical entities such as a production module, a robot, a vehicle, a house, or even a human body \cite{alanis2025subnetworks, adeogun2020towards, berardinelli2021extreme}. These subnetworks would provide capillary wireless coverage, enabling life-critical or mission-critical communications that historically relied on wired connectivity.

In-X deployments promise ultra-low latency, high reliability, and dense spectrum reuse, but also introduce severe challenges in interference management. When many subnetworks operate in close proximity, inter-subnetwork interference becomes a dominant performance-limiting factor. Furthermore, many in-X applications are power-constrained, motivating the need for energy-efficient (EE) operation to extend device lifetime and reduce operational costs. These requirements have motivated recent efforts on interference mitigation via blind repetition, pseudo-random frequency hopping, environment-aware channel allocation, and packet duplication~\cite{adeogun2022enhanced}. Other studies have investigated subband allocation \cite{adeogun2022multi, hakimi2025robust, hakimi2025resilient, adeogun2021learning, du2022multi, madsen2024federated} and power control \cite{abode2024power, li2024power} for supporting in-X subnetworks operation across different scenarios based on both heuristic and machine learning solutions. While these mechanisms demonstrate potential for enhanced performance in terms of spectral efficiency and reliability, they do not explicitly address energy-efficient resource allocation under interference coupling.

Energy efficiency, typically measured in bits-per-Joule, introduces fractional and non-convex optimization objectives that are difficult to handle using classical model-based approaches. At the same time, learning-based methods have recently emerged as powerful tools for wireless resource allocation. In particular, graph neural networks (GNNs) have shown strong potential for modeling interference relationships through message passing and for generalizing across heterogeneous network topologies \cite{gu2023graph, lu2024graph, shen2019graph}. Their structural alignment with wireless interference graphs makes them attractive for scalable and distributed power-control strategies.

Recent work has applied GNNs specifically to industrial in-X subnetworks, demonstrating that interference-aware power control can be learned using only partial channel information, such as subnetwork positions and long-term path gains \cite{abode2024power}. These approaches achieve performance comparable to full-CSI benchmarks while offering robustness to deployment density variations and reduced signaling overhead. More recent studies have leveraged GNNs for joint power and spectrum allocation in interference-limited networks, further highlighting the applicability of message-passing architectures to non-convex, multi-objective wireless optimization problems \cite{marwani2024graph}.

Motivated by these developments, this paper investigates energy-efficient power control for dense, interference-limited in-X subnetworks using a GNN-based architecture. We focus on realistic industrial deployments and consider three optimization formulations that reflect practical system-design perspectives: (i) network-centric EE maximization, (ii) subnetwork-centric EE fairness, and (iii) a multi-objective tradeoff between spectral efficiency (SE) and energy efficiency. 

Our contributions are summarized as follows:
\begin{itemize}
    \item We formulate three complementary EE-driven power-control objectives capturing system-wide efficiency, fairness across subnetworks, and flexible SE--EE tradeoffs.
    \item We study a GNN-based power-control architecture with three different optimization objectives.
    \item We conduct extensive simulations using 3GPP InF channel models and varying deployment densities, demonstrating substantial improvements in both network-level and per-link energy efficiency, while preserving or enhancing spectral efficiency.
\end{itemize}

These results show that GNN-based power control is a viable and scalable approach to enabling energy-aware operation in future 6G in-X subnetworks, supporting the broader vision of replacing wired connections in mission-critical and power-constrained environments.

\section{System Model}

We consider a wireless system composed of $M$ subnetworks. The set of subnetwork indices is denoted $\mathcal{M}=\{1, 2,\cdots, M\}$. Each subnetwork consists of a single access point (AP) serving $K$ user devices. We denote the set of devices in the $m$th subnetwork as $\mathcal{K}_m={1,2,\cdots,K}$. Devices within a subnetwork are assumed to served orthogonally (e.g., via Time Division Multiple Access (TDMA) or Frequency Division Multiple Access (FDMA)), so intra-subnetwork interference is eliminated. Consequently, interference arises only from APs in other subnetworks transmitting on overlapping resources.

The received signal at device $k$ in subnetwork $m$ is
\begin{equation}
y_k^{(m)} = h_{kk}^{(m)} \sqrt{p_k^{(m)}} s_k^{(m)} 
+ \sum_{j \in \mathcal{M}; \, j\neq k} h_{kj}^{(m)} \sqrt{p_j} s_j
+ n_k^{(m)},
\end{equation}
where $s_k^{(m)}$ is the unit-power symbol intended for device $k$, 
$p_k^{(m)} \in [0,P_{\max}]$ is the transmit power allocated from AP $m$ to device $k$, 
$h_{kj}^{(m)}$ denotes the channel gain from AP $j$ (in another subnetwork) to device $k$, 
and $n_k^{(m)} \sim \mathcal{CN}(0,\sigma^2)$ is additive white Gaussian noise.

The signal-to-interference-plus-noise ratio (SINR) at device $k$ in subnetwork $m$ is therefore
\begin{equation}
\gamma_k^{(m)} = \frac{|h_{kk}^{(m)}|^2 p_k^{(m)}}{\sum_{j \in \mathcal{M}; \, j\neq k} |h_{kj}^{(m)}|^2 p_j + \sigma^2},
\end{equation}
and the achievable rate is
\begin{equation}
R_k^{(m)} = \log_2\left(1 + \gamma_k^{(m)}\right).
\end{equation}

\subsection{Power Consumption Model}

The total power consumed to serve device $k$ in subnetwork $m$ is
\begin{equation}
P_{\text{cons},k} = \frac{p_k}{\eta} + P_c,
\end{equation}
where $\eta \in (0,1]$ is the power amplifier efficiency of the AP and $P_c$ is the static circuit power per device. Hardware is assumed homogeneous across subnetworks, and additional overhead (e.g., processing or backhaul) is neglected.

\section{Energy Efficient Power Optimization Problem Formulation}

This paper considers power control for interference-coupled in-X subnetworks under three energy-efficiency-oriented formulations. The formulations represent distinct system-level design philosophies: maximizing the global efficiency of the subnetwork, improving the average subnetwork-level efficiency, and trading off spectral- and energy-efficiency through a scalarized multi-objective criterion. 

\subsection{Network-Centric Energy Efficiency Maximization}

The Network-Centric Energy Efficiency (NCEE) formulation maximizes the global energy efficiency of the entire subnetworks deployment. NCEE is defined as
\begin{equation}
\text{EE}_{\text{net}} = \frac{\sum_{m=1}^M\sum_{k=1}^{K} R_k^{(m)}}{\sum_{m=1}^M\sum_{k=1}^{K} P_{\text{cons},k}^{(m)}}.
\end{equation}

The NCEE, $\text{EE}_{\text{net}}$ captures the number of bits transmitted per Joule consumed across all subnetworks. Maximizing $\text{EE}_{\text{net}}$ encourages the model to jointly adjust power levels to minimize unnecessary power expenditure while preserving aggregate throughput. Links with persistently weak channels may be assigned very low power.

\subsection{Subnetwork-Centric Energy Efficiency Maximization}

The Subnetwork-Centric Energy Efficiency (SCEE) formulation focuses on subnetwork-level efficiency. The energy efficiency of subnetwork $m$ is defined as
\begin{equation}
\text{EE}_m = \frac{\sum_{k=1}^KR_k}{\sum_{k=1}^KP_{\text{cons},k}}.
\end{equation}

The objective is to maximize the average per-subnetwork EE expressed as
\begin{equation}
\text{EE}_{\text{avg}} = \frac{1}{M} \sum_{m=1}^{M} \text{EE}_m.
\end{equation}

Unlike the network-centric objective, this formulation emphasizes fairness: each subnetwork contributes equally to the objective, regardless of its channel conditions. %The GNN is therefore encouraged to balance power allocations more evenly across the subnetwork, which is particularly relevant for industrial or safety-critical scenarios where uniform performance among devices is required.

\subsection{Multi-Objective Spectral- and Energy-Efficiency Tradeoff}

The Multi-Objective spectral- and Energy-Efficiency (MOEE) formulation uses a scalarized objective that jointly accounts for SE and EE. The MOEE is defined as
\begin{equation}
\label{eqMOEE}
J(\mathbf{p}) = \alpha \sum_{m=1}^{M} R_m - (1-\alpha) \sum_{m=1}^{M} P_{\text{cons},m}, \quad \alpha \in [0,1].
\end{equation}

The formulation in \eqref{eqMOEE} allows the system designer to interpolate between SE-centric operation ($\alpha \to 1$) and EE-centric operation ($\alpha \to 0$). Although simpler than the fractional EE objectives, it retains practical interpretability and provides a tunable interface for deployment-specific requirements. This tradeoff is useful when throughput demands vary across devices or over time, enabling operation along the SE–EE frontier rather than at a fixed efficiency target. Note that the MOEE formulation in \eqref{eqMOEE}
corresponds to that standard Power Control GNN (PCGNN) proposed in \cite{abode2024power} when $\alpha = 1$. 

The three formulations presented in this section enable a systematic comparison of how objective design influences learned power-control behavior. While NCEE emphasizes overall system efficiency, SCEE objective promotes fairness, and MOEE provides a mechanism to control the tradeoff between spectral efficiency and energy expenditure. Since all three objectives are differentiable with respect to the transmit powers, they can be used directly as training losses in the GNN framework described next.

%\section{Graph Based Learning for Energy Efficient Power Control}
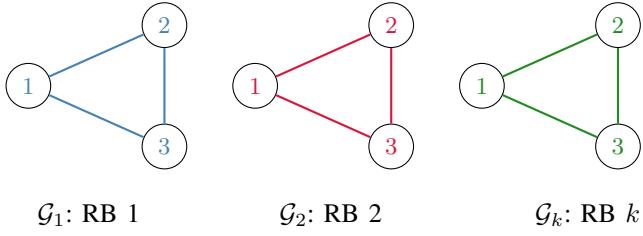
\begin{figure}[t]
\centering
\begin{tikzpicture}[>=stealth, node distance=1.8cm]

% Colors for different graphs
\definecolor{g1}{RGB}{70,130,180}
\definecolor{g2}{RGB}{220,20,60}
\definecolor{g3}{RGB}{34,139,34}

% Node style: colored text, no shape, no border
\tikzset{
    devnode/.style={circle,font=\small, draw, text=#1}
}

% --- Graph G1 (Resource block 1)
\node[devnode=g1] (g11) at (-1.8,0) {$1$};
\node[devnode=g1] (g12) at (0,0.8) {$2$};
\node[devnode=g1] (g13) at (0,-0.8) {$3$};

\draw[g1, thick] (g11) -- (g12);
\draw[g1, thick] (g11) -- (g13);
\draw[g1, thick] (g12) -- (g13);

\node at (-1.0,-1.7) {$\mathcal{G}_1$: RB 1};

% --- Graph G2 (Resource block 2)
\node[devnode=g2] (g21) at (1.2,0) {$1$};
\node[devnode=g2] (g22) at (3.0,0.8) {$2$};
\node[devnode=g2] (g23) at (3.0,-0.8) {$3$};

\draw[g2, thick] (g21) -- (g22);
\draw[g2, thick] (g21) -- (g23);
\draw[g2, thick] (g22) -- (g23);

\node at (2.2,-1.7) {$\mathcal{G}_2$: RB 2};

% --- Graph G3 (Resource block k)
\node[devnode=g3] (g31) at (4.2,0) {$1$};
\node[devnode=g3] (g32) at (6.0,0.8) {$2$};
\node[devnode=g3] (g33) at (6.0,-0.8) {$3$};

\draw[g3, thick] (g31) -- (g32);
\draw[g3, thick] (g31) -- (g33);
\draw[g3, thick] (g32) -- (g33);

\node at (5.6,-1.7) {$\mathcal{G}_k$: RB $k$};

% Titles
% \node at (0.9,1.6) {\footnotesize One device per subnetwork};
% \node at (0.9,1.3) {\footnotesize (all share RB 1)};
% \node at (5.9,1.6) {\footnotesize One device per subnetwork};
% \node at (5.9,1.3) {\footnotesize (all share RB 2)};
% \node at (10.9,1.6) {\footnotesize One device per subnetwork};
% \node at (10.9,1.3) {\footnotesize (all share RB $k$)};

\end{tikzpicture}

\caption{Decomposition into $K$ independent interference graphs.
Each graph $\mathcal{G}_k$ contains one device per subnetwork using resource block $k$, forming a fully connected interference graph.}
\label{fig:multigraph}
\end{figure}

\begin{table}[t]
\centering
\caption{Simulation Parameters}
\begin{tabular}{l l}
\toprule
\textbf{Parameter} & \textbf{Value} \\
\midrule
Indoor Factory size & $10$m $\times 10$m\\
Number of subnetworks & $M = 10$ \\
Devices per subnetwork & $K = 1$ \\
Maximum transmit power & $P_{\max} = 1$ W (normalized) \\
Bandwidth & $5$ MHz \\
Power amplifier efficiency & $\eta = 0.8$ \\
Circuit power per device & $P_c = 0.1$ W \\
Channel model & 3GPP InF \\
Pathloss exponent & $2.1$\\
Shadowing standard deviation [dB] & $7$\\
Training samples & $50{,}000$  \\
Batch size & $64$ \\
GNN layers & $3$ message-passing layers \\
Hidden units per layer & $128$ \\
Activation function & ReLU \\
Optimizer & Adam \\
Learning rate & $10^{-3}$ \\
Number of training epochs & $200$ \\
%Loss functions & Network-wide EE, Average per-link EE, Multi-objective EE-SE \\
%Evaluation metrics & EE (bit/Joule), Sum-rate (bits/s/Hz), Power allocation \\
\bottomrule
\end{tabular}
\label{tab:sim_params}
\end{table}

\begin{figure*}[]
    \centering
    \subfloat[Average network EE.]{
        \includegraphics[width=0.48\linewidth]{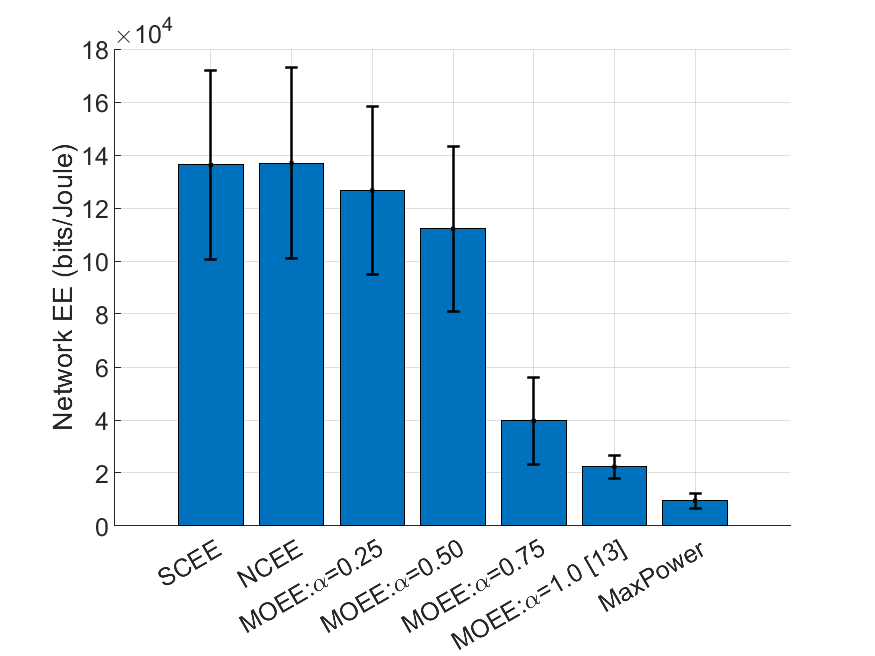}
        \label{fig:avNEE}
    }
    \hfill
    \subfloat[Average spectral efficiency.]{
        \includegraphics[width=0.48\linewidth]{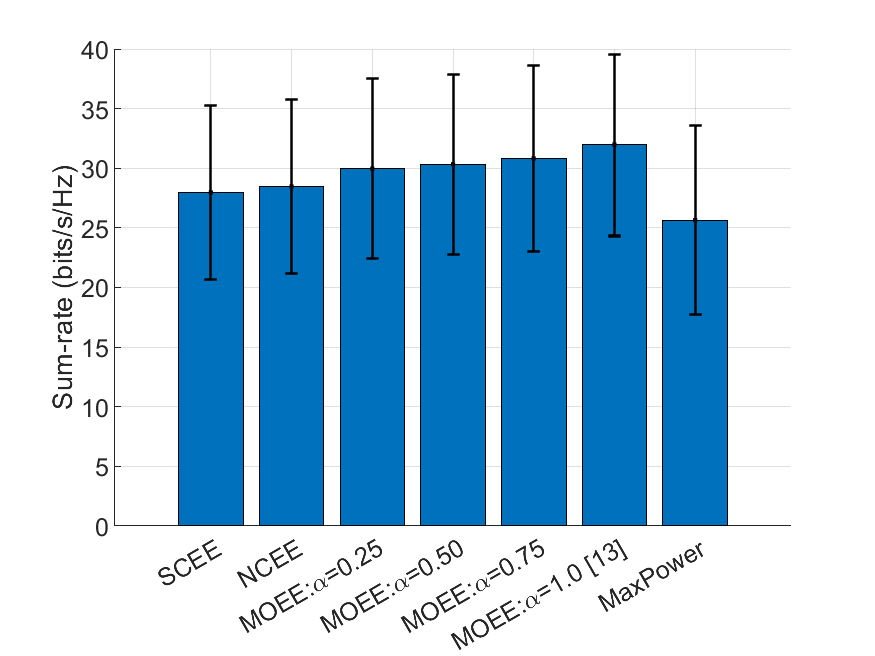}
        \label{fig:avgSE}
    }
    \caption{Spectral efficiency and network energy efficiency performance for the different formulations with $M=10$ subnetworks.Error bars denote $\pm$ standard deviation across test snapshots.}
    \label{fig:boxplots}
\end{figure*}

\begin{figure*}[]
    \centering
    \subfloat[Network EE.]{
        \includegraphics[width=0.48\linewidth]{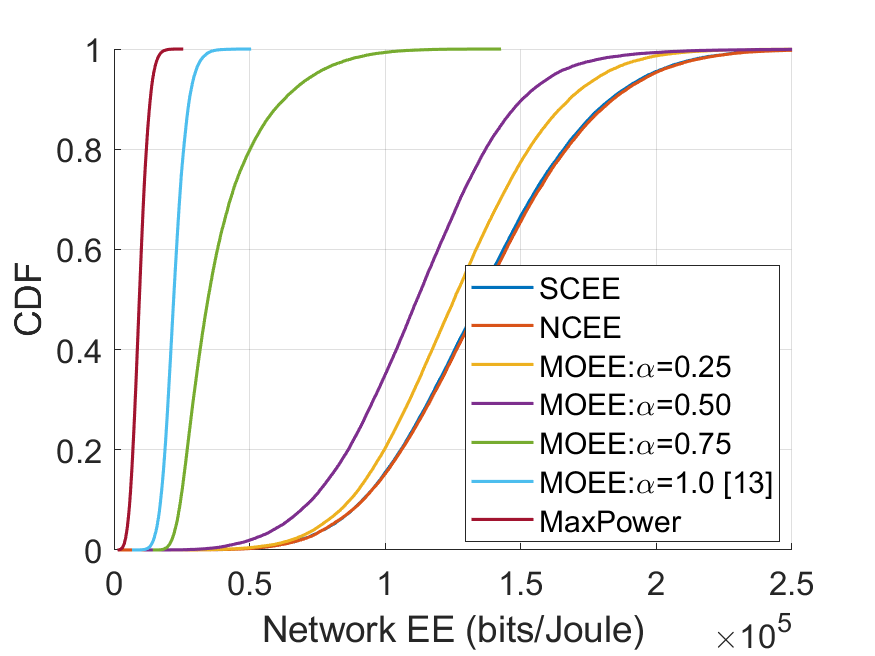}
        \label{fig:NEECDF}
    }
    \hfill
    \subfloat[Per link EE.]{
        \includegraphics[width=0.48\linewidth]{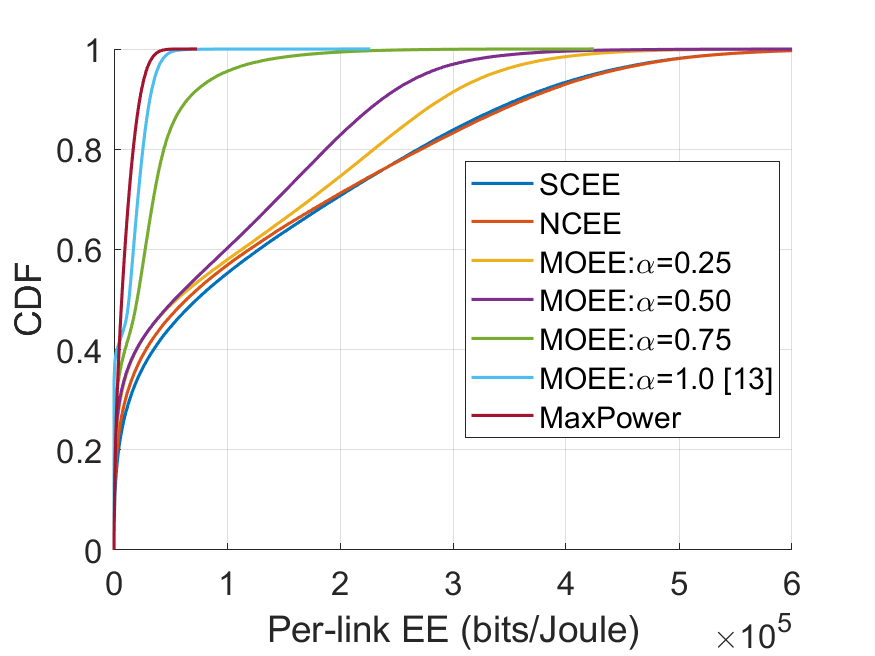}
        \label{fig:PLEE}
    }
    \hfill
    \subfloat[Sum SE.]{
        \includegraphics[width=0.48\linewidth]{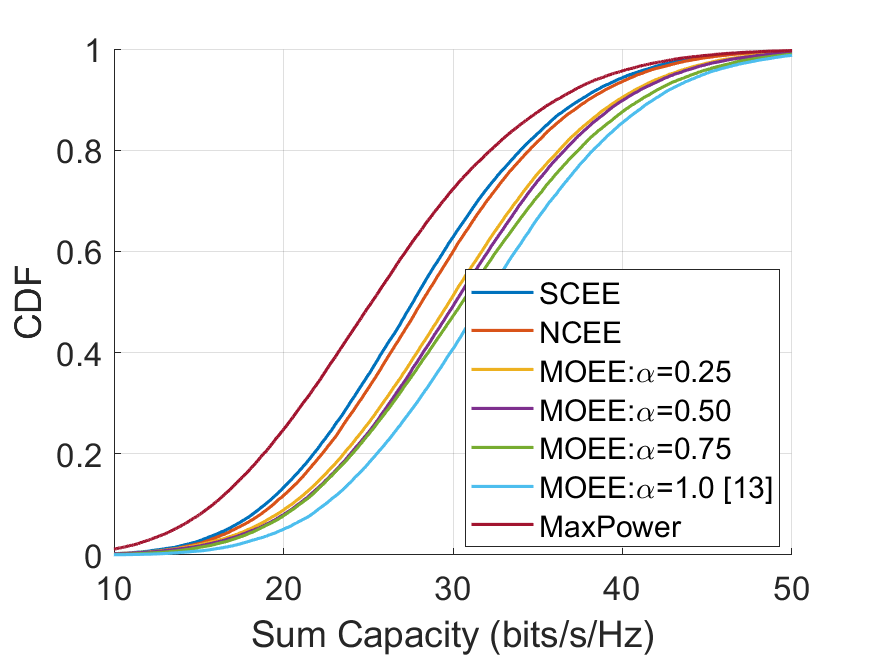}
        \label{fig:SE}
    }
    \hfill
    \subfloat[Per link SE.]{
        \includegraphics[width=0.48\linewidth]{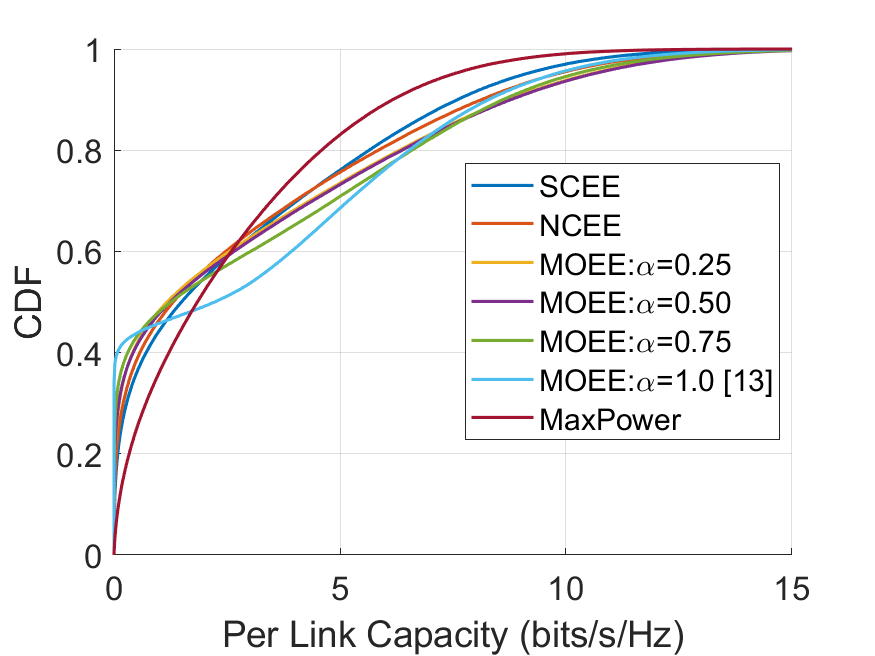}
        \label{fig:avSE}
    }
    \caption{CDF plots of EE and SE with $M=10$ subnetworks.}
    \label{fig:fig3}
\end{figure*}

\begin{figure}[t!]
    \centering
    \includegraphics[width=0.98\linewidth]{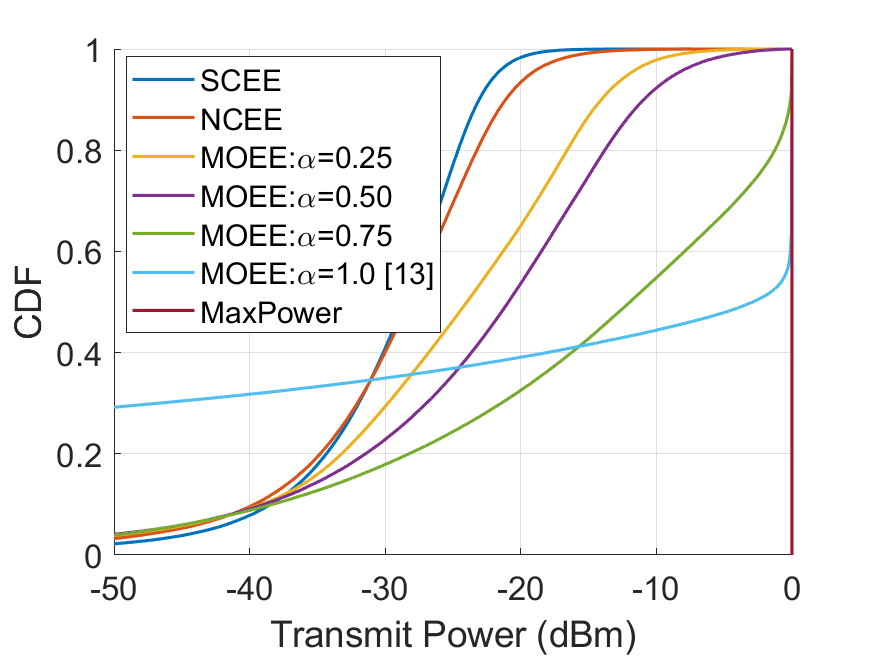}
    \caption{CDF of transmit power allocation.}
    \label{fig:txPow}
\end{figure}
\begin{figure}[t!]
    \centering
    \includegraphics[width=0.98\linewidth]{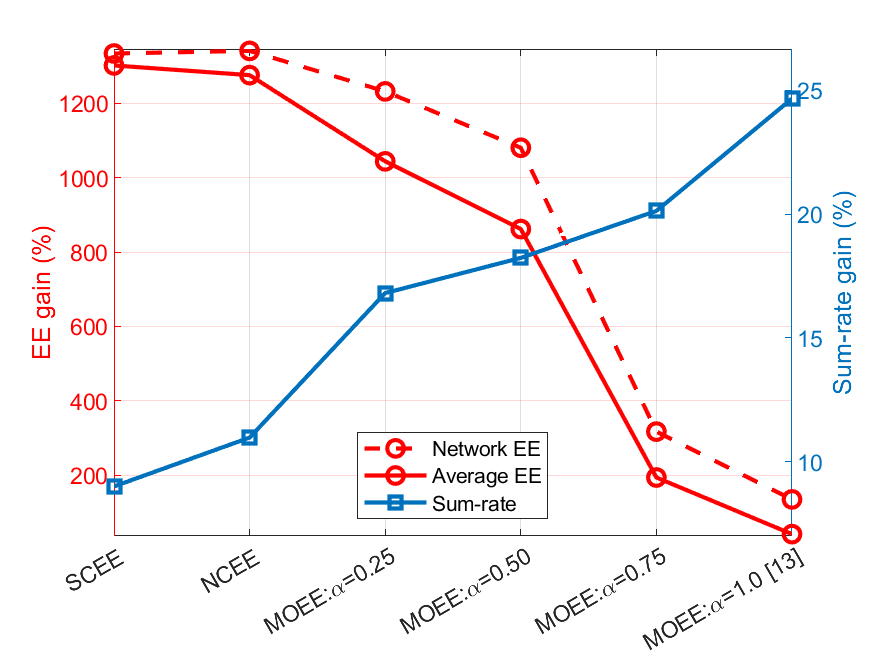}
    \caption{EE versus SE performance tradeoff.}
    \label{fig:EEvSE}
\end{figure}
\section{GNN for Energy Efficient Power Control}

We adopt a graph neural network (GNN) framework specifically tailored to the structure of in-X subnetworks with orthogonal intra-subnetwork transmissions. Because only devices sharing the same resource block interfere with each other, the overall network decomposes into $K$ independent interference graphs. This decomposition significantly reduces the dimensionality of the learning problem while allowing the model to exploit the structured interference pattern inherent to the system.

\subsection{Graph Representation}

Each subnetwork allocates orthogonal transmission resources to its $K$ devices, eliminating intra-subnetwork interference. Devices occupying the same resource index $k$ across different subnetworks transmit simultaneously on the same spectrum band, and therefore form an interference group. This enables the full system to be expressed as $K$ independent graphs:
\[
\mathcal{G}_k = (\mathcal{V}_k, \mathcal{E}_k), \qquad k = 1,\ldots,K.
\]

For a fixed resource index $k$, the node set is
\[
\mathcal{V}_k = \{1,2,\ldots,M\},
\]
where node $m$ represents the $k$-th device in subnetwork $m$. Since all such devices share the same time–frequency resource, they mutually interfere. Thus, $\mathcal{G}_k$ is modeled as a fully connected directed graph as illustrated in Fig.~\ref{fig:multigraph}.

Each node in $\mathcal{G}_k$ carries a feature vector containing locally available information such as desired-link channel gain, device position, and normalized power budget. Edge features encode the cross-subnetwork interference channels $|h_{j,i}|^2$, derived from path loss, shadowing, and small-scale fading. A global feature vector stores system-level parameters including noise power $\sigma^2$, bandwidth, PA efficiency $\eta$, and circuit power $P_c$.

This multi-graph representation aligns with the physical structure of the system and significantly improves scalability. The GNN processes all $K$ graphs using shared parameters, allowing the resulting policy to generalize across different subnetwork densities.

\subsection{Message Passing and Feature Aggregation}

For each interference graph $\mathcal{G}_k$, the GNN applies $L$ rounds of message passing. Nodes exchange messages derived from their hidden states and the edge features representing cross-subnetwork interference.

At message-passing layer $\ell$, node $m$ receives messages from all other nodes $j\neq m$:
\[
m_{m\leftarrow j}^{(\ell)} = 
\psi\!\left( h_j^{(\ell)}, e_{j,m} \right),
\]
where $h_j^{(\ell)}$ is the hidden state of node $j$ at layer $\ell$, and $e_{j,m}$ encodes the interfering channel from subnetwork $j$ to $m$.

Messages are aggregated using a permutation-invariant operator:
\[
\bar{m}_m^{(\ell)} = 
\sum_{j\in\mathcal{V}_k,\, j\neq m} 
m_{m\leftarrow j}^{(\ell)}.
\]

The hidden state is then updated as
\[
h_m^{(\ell+1)} = 
\phi\!\left(
h_m^{(\ell)},
\bar{m}_m^{(\ell)},
\mathbf{u}
\right),
\]
where $\mathbf{u}$ denotes the global feature vector.  
The functions $\psi$ and $\phi$ are implemented as multilayer perceptrons, allowing the network to learn the structure of interference interactions across subnetworks. This mechanism ensures that the learned representations reflect the collective interference impact of devices sharing the same resource block, while maintaining permutation invariance and size generalization properties.

\subsection{Power Allocation}

After $L$ message-passing layers, the hidden representation of each node is mapped to a raw power prediction:
\[
\tilde{p}_m^{(k)} = 
f_{\mathrm{out}}\!\left( h_m^{(L)} \right).
\]

To enforce the physical constraint $0 \le p_m^{(k)} \le P_{\max}$, the output is normalized using a sigmoid function, $\sigma$:
\[
p_m^{(k)} = 
P_{\max}\,\sigma\!\left(\tilde{p}_m^{(k)}\right).
\]

This approach is fully differentiable and avoids the need for explicit projection or clipping operations during training.

\subsection{Training for Energy-Efficiency Objectives}

The GNN parameters are trained end-to-end using the differentiable objectives defined in Section~III, applied independently on the $K$ graphs. Specifically, we train separate models using each of the following loss functions:

\begin{align}
\mathcal{L}_{\text{ncee}}(\mathbf{p})
&= - \frac{\sum_{m=1}^M R_k^{(m)}}{\sum_{m=1}^M P_{\text{cons},k}^{(m)}}\quad \quad \forall k, 
\label{eq:loss_netEE} \\[4pt]
\mathcal{L}_{\text{scee}}(\mathbf{p})
&= - \frac{1}{M}\sum_{m=1}^M\frac{ R_k^{(m)}}{ P_{\text{cons},k}^{(m)}}\quad \quad \forall k,
\label{eq:loss_avgEE} \\[4pt]
\mathcal{L}_{\text{moee}}(\mathbf{p})
&= - \left( 
\alpha \sum_{m=1}^M R_k^{(m)}
- (1-\alpha) \sum_{m=1}^{M} P_{\text{cons},k}^{(m)}
\right), \quad \quad \forall k.
\label{eq:loss_mo}
\end{align}

\subsection{Scalability and Generalization}

The decomposition into $K$ fixed-size interference graphs ensures linear scaling with the number of devices per subnetwork and enables strong generalization across deployment scenarios. Since message passing relies exclusively on locally aggregated interference information, the learned power control policy naturally adapts to variations in subnetwork density, shadowing realizations, network topology, and heterogeneous device placements. This robustness to environmental and structural changes is consistent with the generalization properties observed in the subnetwork-based GNN power control framework in \cite{abode2025goal}, where a single trained model was shown to maintain performance across diverse spatial layouts and propagation conditions.

\section{Performance Evaluation}

\subsection{Simulation Settings}

We consider a dense deployment of $10$ in-X subnetwork in a $10 \times 10$~m$^2$ indoor factory environment. Without loss of generality, we assume that each subnetwork consists of a single device. Channels are modeled using 3GPP models \cite{3gpp38901} for in-factory environments with path-loss exponent $\gamma = 2.1$ and shadowing standard deviation $7$~dB. The system bandwidth is $5$~MHz, the power amplifier efficiency is $\eta = 0.8$, and each device incurs a static circuit power of $P_c = 0.1$~W.

The GNN architecture consists of three message-passing layers with $128$ hidden units each and ReLU activations. Models are trained using the Adam optimizer with a learning rate of $10^{-3}$, a batch size of $64$, and $500$ training epochs. Each scenario is trained on $50{,}000$ samples. Other simulation parameters are defined in Table~\ref{tab:sim_params}. 

\subsection{Performance Results}
We now present a comparative performance evaluation of the three formulations. 

Fig.~\ref{fig:avNEE} and Fig.~\ref{fig:avgSE} present average network energy efficiency (EE) and average spectral efficiency (SE) under the three objective formulations: NCEE, SCEE, and MOEE with varying trade-off parameter, $\alpha$, respectively. The NCEE objective produces the highest network EE while the SCEE objective yields marginally lower network EE. The MOEE curve traces intermediate operating points: as $\alpha$ increases (i.e., more SE emphasis) average SE grows and EE falls, and vice-versa. Across the tested regimes, the learned GNN policies substantially outperform the maximum transmit power benchmark. 

Fig.~\ref{fig:fig3} shows the empirical CDF of network-level energy efficiency across testing snapshots for the three formulations.  The figure shows that maximizing network centric EE in NCEE translates to the highest network EE across the entire distribution. The CDF for NCEE is strongly shifted to the right relative to other formulations, demonstrating that the NCEE-trained GNN consistently identifies power allocations that extract near-optimal EE performance under varying interference conditions.This is expected due to the direct optimization of network EE in NCEE objective. The figure also shows that distribution of network EE for MOEE depends on the weight parameter. When $\alpha$ is tuned to $1$ (favouring SE), the CDF shifts leftwards due to increased power expenditure; when $\alpha$ favors EE, the curve moves closer to NCEE. This behavior reflects the tradeoff inherent in the MOEE objective, where gains in spectral efficiency come at the cost of reduced energy efficiency.

Fig.~\ref{fig:EEvSE} shows the achievable energy–spectral efficiency frontier for NCEE, SCEE and the MOEE objective as the tradeoff parameter $\alpha$ varies. By excluding the points for NCEE and SCEE, the resulting curve approximates a Pareto frontier: low $\alpha$ achieves high EE at modest SE, high $\alpha$ attains high SE at reduced EE, and a pronounced \emph{knee} appears around $\alpha =0.5$ where small SE increases require large EE sacrifices. This knee marks attractive operating points for practical deployments that need a balanced throughput and energy savings performance. %The shape of the frontier indicates that moderate $\alpha$ values (mid-range weighting) provide sizeable SE gains for relatively small EE loss, giving designers a compact knob to trade network objectives.

\section{Conclusion}
This paper studied energy-efficient power control in 6G in-X subnetworks. We considered a graph neural network (GNN) framework that captures inter-subnetwork interference and learns distributed power allocation policies. Three formulations were considered: network-centric, subnetwork-centric, and a multi-objective approach balancing energy efficiency and sum-rate. Simulation results demonstrate that GNN-based solutions can achieve substantial energy efficiency improvements while maintaining high sum-rate performance. Specifically, network energy efficiency gains of up to 1341\%, average per-device energy efficiency improvements of up to 1302\%, and sum-rate gains of up to 24.7\% were observed relative to a maximum power baseline, depending on the optimization objective and trade-off weights. These results highlight the effectiveness and flexibility of learning-based approaches for interference-aware power control in future 6G subnetworks. 

%The study provides a unified framework for evaluating trade-offs between global network performance and per-subnetwork energy efficiency, offering practical insights for energy-aware resource allocation in next-generation wireless systems.

\bibliographystyle{IEEEtran}
\bibliography{ref.bib}
\end{document}